\documentclass[12pt]{iopart}
\usepackage{iopams}
\usepackage{setstack}

\usepackage{ae}
\usepackage{color}
\usepackage{url}

\usepackage{graphicx}

\DeclareGraphicsExtensions{.eps}
\graphicspath{{figures/}}

\newcommand{\Zeff}{Z$_{\mathrm{eff}}$ }
\newcommand{\Zeffd}{Z$_{\mathrm{eff}}$.}
\newcommand{\ud}{\mathrm{d}}

\begin{document}

\title{The effect of ITER-like wall on runaway electron generation in JET}

\author{G. Papp$^{1,2}$, T. F\"ul\"op$^{1}$, T. Feh\'er$^{3}$, P. C. de Vries$^{4}$, V. Riccardo$^{5}$, C. Reux$^{6}$, M. Lehnen$^{7}$, V. Kiptily$^{5}$, V. V. Plyusnin$^{8}$, B. Alper$^{5}$ and JET EFDA contributors$^{9}$}
\address{\small JET-EFDA Culham Science Centre, OX15 3DB, Abingdon, UK\\
  \small $^{1}$Department of Applied Physics, Nuclear Engineering, Chalmers University of Technology and Euratom-VR Association, SE-41296 G\"oteborg, Sweden\\
  \small $^{2}$Department of Nuclear Techniques, Budapest University of Technology and Economics, Association EURATOM, H-1111 Budapest, Hungary\\
  \small $^{3}$Max Planck Institute for Plasma Physics, EURATOM Association, Boltzmannstr. 2, 85748 Garching, Germany\\
  \small $^{4}$FOM institute DIFFER, Association EURATOM-FOM, PO Box 120, Nieuwegein, Netherlands\\
  \small $^{5}$CCFE/EURATOM Association, Culham Science Centre, Abingdon, OX14~3DB, UK\\
  \small $^{6}$Ecole Polytechnique, CNRS, 91128 Palaiseau Cedex, France\\
  \small $^{7}$Institut f\"ur Energie- und Klimatforschung-IEK-4, Forschungszentrum J\"ulich GmbH, EURATOM Association, 52425 J\"ulich, Germany\\
  \small $^{8}$Instituto de Plasmas e Fus$\tilde{\mathrm{a}}$o Nuclear/IST, Associacao EURATOM-IST, Av.~Rovisco Pais, 1049-001 Lisbon, Portugal\\
  \small $^{9}$See the Appendix of F. Romanelli et al., Proceedings of the 24th IAEA Fusion Energy Conference 2012, San Diego, US\\
} \ead{papp@chalmers.se}

\begin{abstract}

This paper investigates the effect of the ITER-like wall (ILW) on runaway electron (RE) generation through a comparative study of similar slow argon injection JET disruptions, performed with different wall materials. In the carbon wall case, a runaway electron plateau is observed, while in the ITER-like wall case, the current quench is slower and the runaway current is negligibly small.
The aim of the paper is to shed light on the reason for these differences by detailed numerical modelling to study which factors affected the RE formation.
The post-disruption current profile is calculated by a one-dimensional model of electric
  field, temperature and runaway current taking into account the
  impurity injection. Scans of various impurity contents are performed
  and agreement with the experimental scenarios is obtained for
  reasonable argon- and wall impurity contents. Our modelling shows that the reason for
  the changed RE dynamics is a complex, combined effect of the differences in plasma parameter profiles, the radiation
  characteristics of beryllium and carbon, and the difference of the
  injected argon amount. These together lead to a significantly higher Dreicer
  generation rate in the carbon wall case, which is less prone to be suppressed by RE loss mechanisms.
The results indicate that the differences are greatly reduced above $\sim$50\% argon content, suggesting that significant RE current is expected in future massive gas injection experiments on both JET and ITER.

\end{abstract}



\section{Introduction}
Runaway electrons (RE) with energies of several megaelectronvolts have
been observed during disruptions in JET
\cite{wesson89disruptions,gill93generation,gill02behaviour,riccardo03disruptions,plyusnin06study} and other large
tokamaks.
These intense electron beams are the result of the sudden cooling in
connection with disruptions and may cause severe damage to the plasma
facing components and vacuum vessel wall. Therefore they are
considered to be a potential threat to the operation of tokamaks with
high currents, such as ITER \cite{hender07iter}. Extra care is necessary when operating with easy to melt materials, such as beryllium \cite{bazylev13modeling}.
Understanding of the dynamics of these RE
beams could help in developing methods for avoiding the beam formation
or at least localized damage to the wall.  Several tokamaks have
studied the behaviour of these electrons during unintentional or
deliberate disruptions caused for example by an intense argon or other noble
gas puff \cite{wesson89disruptions,gill93generation,lehnen11disruption}. However, proper theoretical
understanding of the differences in the behaviour of
runaways is still missing.  One of the open questions is the different
runaway behaviour in the presence of carbon and beryllium wall impurities, a
question which recently gained interest in the view of the new
ITER-like wall (ILW) installed at JET. The ILW comprises solid beryllium limiters and a combination of bulk tungsten and tungsten-coated carbon fibre composite divertor tiles \cite{matthews11ilw}.

The ILW has a significant impact on disruption physics in general
\cite{devries12impact,lehnen13disruption}. One of the major differences compared to disruptions
with the carbon wall is that a lower fraction of energy is radiated
during the disruption process, yielding higher plasma temperatures
after the thermal quench. This will in turn affect the current quench
times, and also the runaway beam formation. It has been observed that
a slower current quench reduces the runaway generation. 
Drawing experimental conclusions at present time is difficult due to the limited number of runaway experiments carried out with the ILW so far.
Modelling is required in order to understand the role of the different wall in the runaway behaviour and to aid the upcoming extensive runaway experiments.
The aim of this paper is to perform a comparative study of two similar
L-mode limiter discharges, performed with different wall materials and to provide a deeper insight in the differences. In both cases the
disruption was induced by slow argon injection. We will calculate the
post-disruption current profile from the plasma parameters in the two
specific JET disruptive discharges, using simulations based on a
one-dimensional model that solves coupled differential equations for
the runaway density, heat diffusion and the plasma current in the
presence of impurity injection \cite{feher11simulation}.

The structure of the paper is the following. In
sec.~\ref{sec:scenarios} we describe the scenarios that has been
selected to represent typical (but similar) disruptions in JET in the
presence of the carbon and beryllium wall, respectively. Here we also
summarize the plasma parameters used in the simulations and the
experimentally observed quantities, such as runaway currents,
thermal- and current quench times. Section
\ref{sec:model} is devoted to the description of the numerical model
we use for studying the runaway electron dynamics. In sec.~\ref{sec:results} we
present the results of the simulations. First we describe simulations
where the temperature evolution is taken from the experiment (without
modelling the impurity radiation and ionization process). Next, we
include the effect of impurity injection and perform scans over the
argon and background impurity (carbon or beryllium) contents that
cannot be accurately determined experimentally. We calculate the
plasma current evolution and radiation power and compare the
simulations with experimental observations in the two cases (CFC vs ILW). Finally
we assess the effect of magnetic perturbations and determine what
level of perturbation is needed for runaway beam suppression. The
conclusions are presented and discussed in sec.~\ref{sec:conclusion}.

\section{Scenarios selected for modelling} \label{sec:scenarios}

The two discharges selected to represent typical triggered disruption
and runaway behaviour in JET are \#79423 (carbon wall case) and \#81928
(ILW case).  These were {L-mode} limiter discharges, intended to be as similar as possible
with respect to plasma parameters and triggering of the disruption.
The basic plasma parameters are shown in table \ref{tab:parameters}.

\begin{table}
\caption{\label{tab:parameters}The pre-disruption plasma parameters used in the simulations.}
\begin{indented}
\item[]
\begin{tabular}{@{}llll}
\br
Parameter name&Notation&\#79423 (CFC)&\#81928 (ILW)\\
\mr
Major radius & $R_0$ & 3 m & 3 m\\
Minor radius & $a$ & 0.88 m & 0.86 m\\
Magnetic field on axis & $B$ & 2 T & 2 T\\
Plasma current & $I_p$ & 1.93 MA & 1.89 MA\\
Elongation & $\kappa$ &1.3 & 1.3\\
Effective charge & $Z_\mathrm{eff}$ & 2.2 $\pm$ 20\% & 2.5 $\pm$ 20\% \\
Density on axis & $n_0$ & $2.59\cdot 10^{19}$ m$^{-3}$ &  $3.17\cdot 10^{19}$ m$^{-3}$\\
Density profile & $n_e(r)$ & $n_0 (1-1.27 \cdot r^2)^{0.43}$ &$n_0 (1-1.32 \cdot r^2)^{0.4}$ \\
Temperature on axis & $T_0$ & 2.17 keV & 2.45 keV\\
Temperature profile & $T_e(r)$ & $T_0 (1-1.03 \cdot r^2)^2$ & $T_0 (1-0.98 \cdot r^2)^2$\\
Coulomb logarithm (on axis) & $\ln \Lambda$ & 23.2 & 22.7 \\
$q$ on axis & $q_0$ & 1.03 & 0.95\\
$q$ on edge & $q_a$ & 4.05 & 4 \\
 $q_{95}$& $q_{95}$ & 3.56 & 3.5 \\ 
$q$ profile & $q(r/a)$ & $\alpha=1.11$  & $\alpha=1.16$\\
$~~q_0\left(1-[1-(q_0/q_a)^{1/\alpha}]\cdot (r/a)^2\right)^{-\alpha}$ &&&\\
\br
\end{tabular}
\end{indented}
\end{table}

\begin{figure}[htb!]
\begin{center}
\includegraphics[height=55mm]{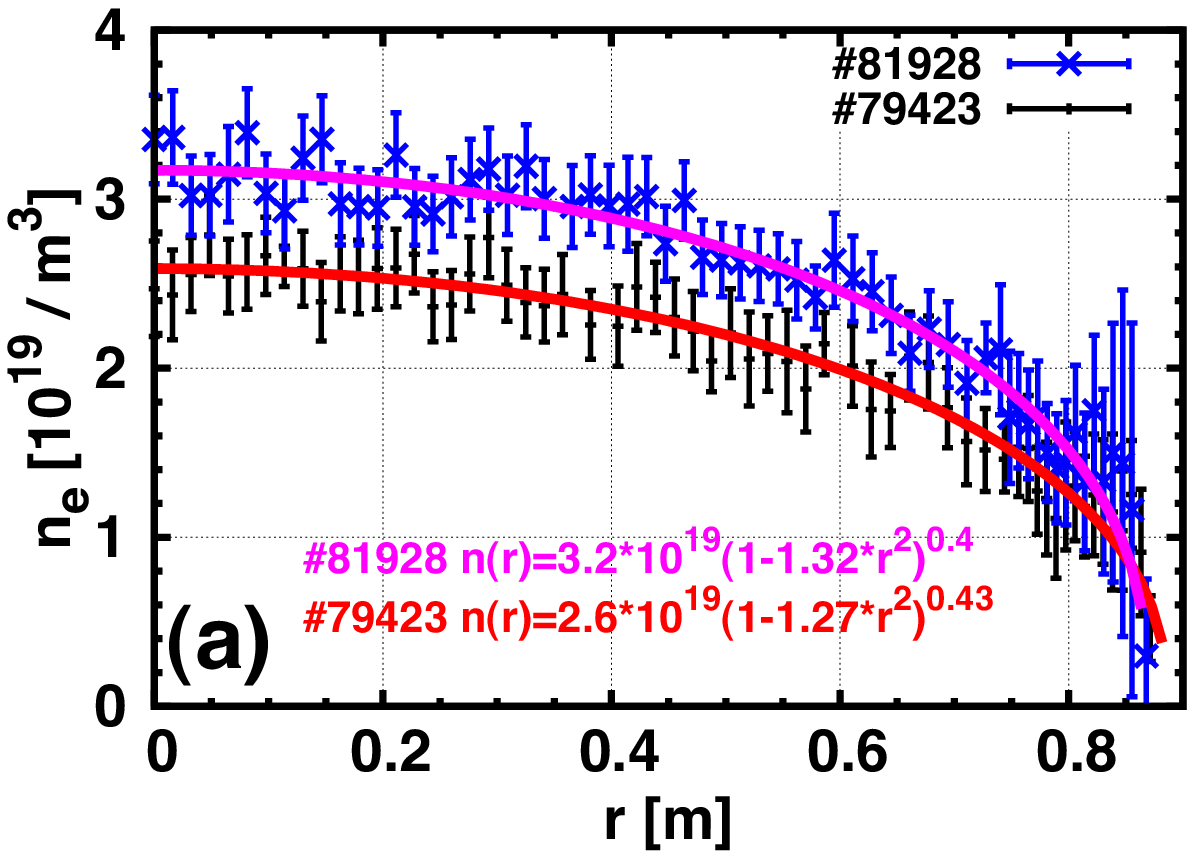}\hfill
\includegraphics[height=55mm]{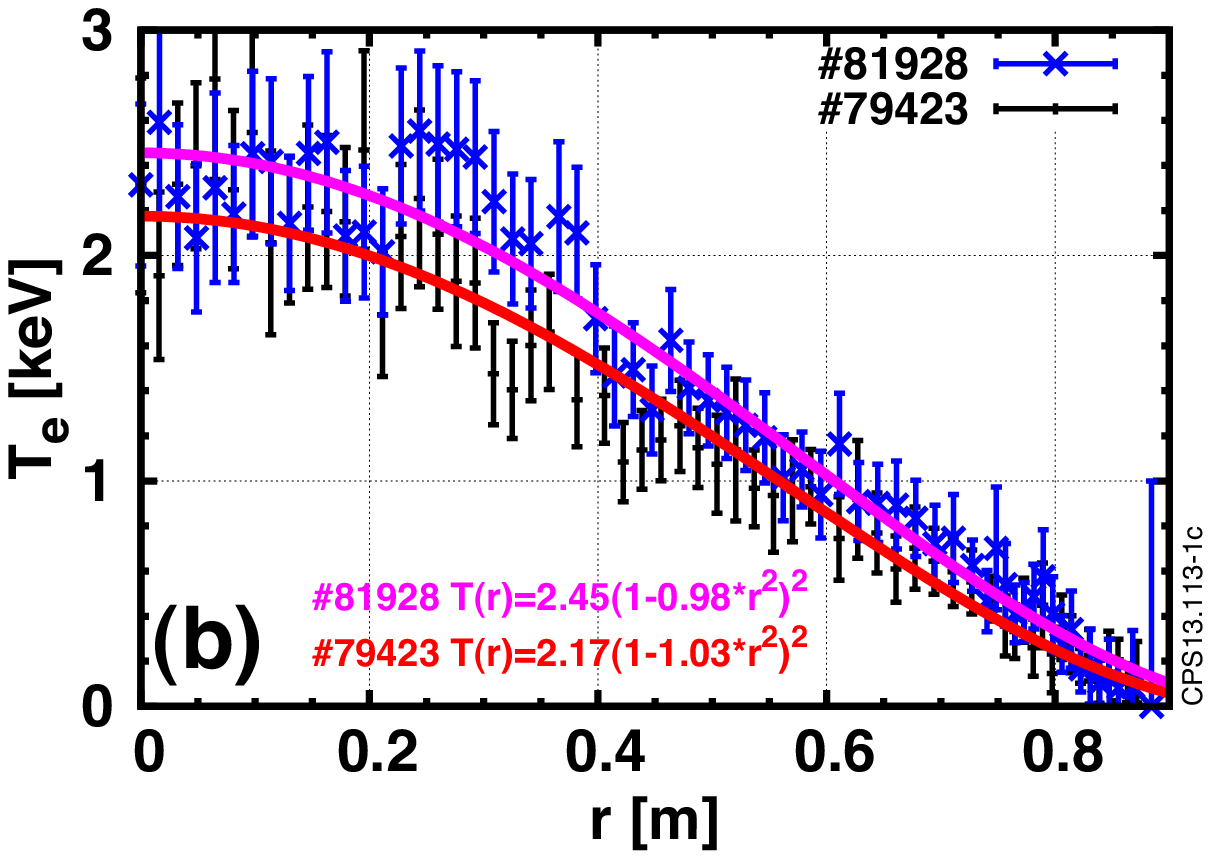}\\
\caption{\label{fig:paramsnete}{Typical (a) electron density- (b)
    electron temperature profiles for the two discharges recorded by Thomson scattering right before the argon valve trigger.}}
\end{center}
\end{figure}
The pre-disruption density- and
temperature profiles are shown in fig \ref{fig:paramsnete}.
The temperature and density profiles were obtained from
Thomson scattering, right before the argon gas valve was triggered. For easier implementation we
fitted typical parameter profile shape functions in the form of
$A_0(1-b\cdot r^2)^c$ to the density and temperature datasets and used
the fits as the simulation input (see table \ref{tab:parameters}). The fitted parameters are shown with two decimal precision. Figures
\ref{fig:paramsnete}a-b contain the experimental uncertainty of the data
which is also taken into account in the fits.



\begin{figure}[htb!]
\begin{center}
\includegraphics[width=0.49\linewidth]{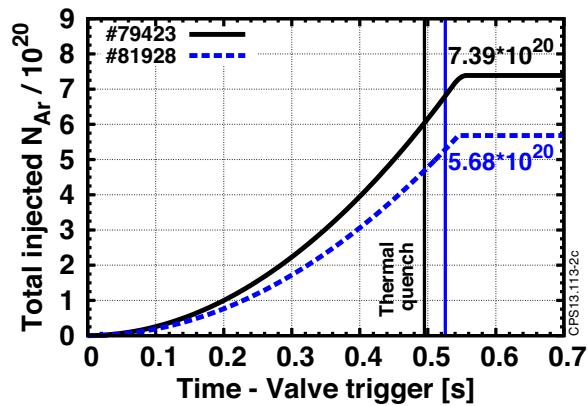}
\caption{\label{fig:inject}{Total number of injected Ar atoms as a function of time. The time
axis is shifted to the valve trigger time.}}
\end{center}
\end{figure}

The disruption was triggered with a slow, controlled injection of room
temperature neutral argon with a linearly increasing injection rate. The valves were triggered
at 21.5~s (\#79423) and 20~s (\#81928). The total number of injected
argon atoms was $7.39\cdot 10^{20}$ for \#79423, 30\% higher than the $5.68 \cdot 10^{20}$ amount
for \#81928, as is shown in figure \ref{fig:inject}.
The total amount of injected impurities at the time of the
thermal quench is only different by $\sim$18\% as indicated by the
vertical lines in figure \ref{fig:inject}.  Note that, this stands
for the {\em injected} material amount, the assimilation rate for the
two cases is unknown.

\begin{figure}[htb!]
\begin{center}
\includegraphics[height=51mm]{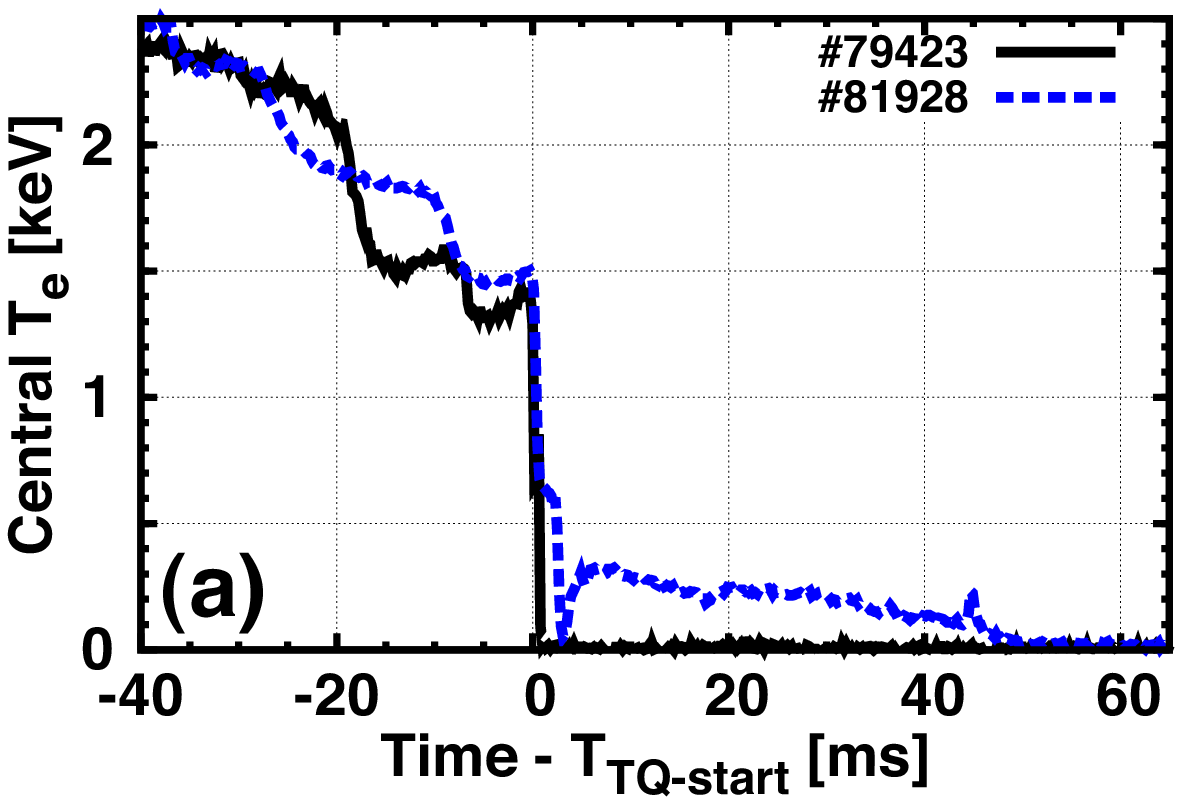}\hfill
\includegraphics[height=51mm]{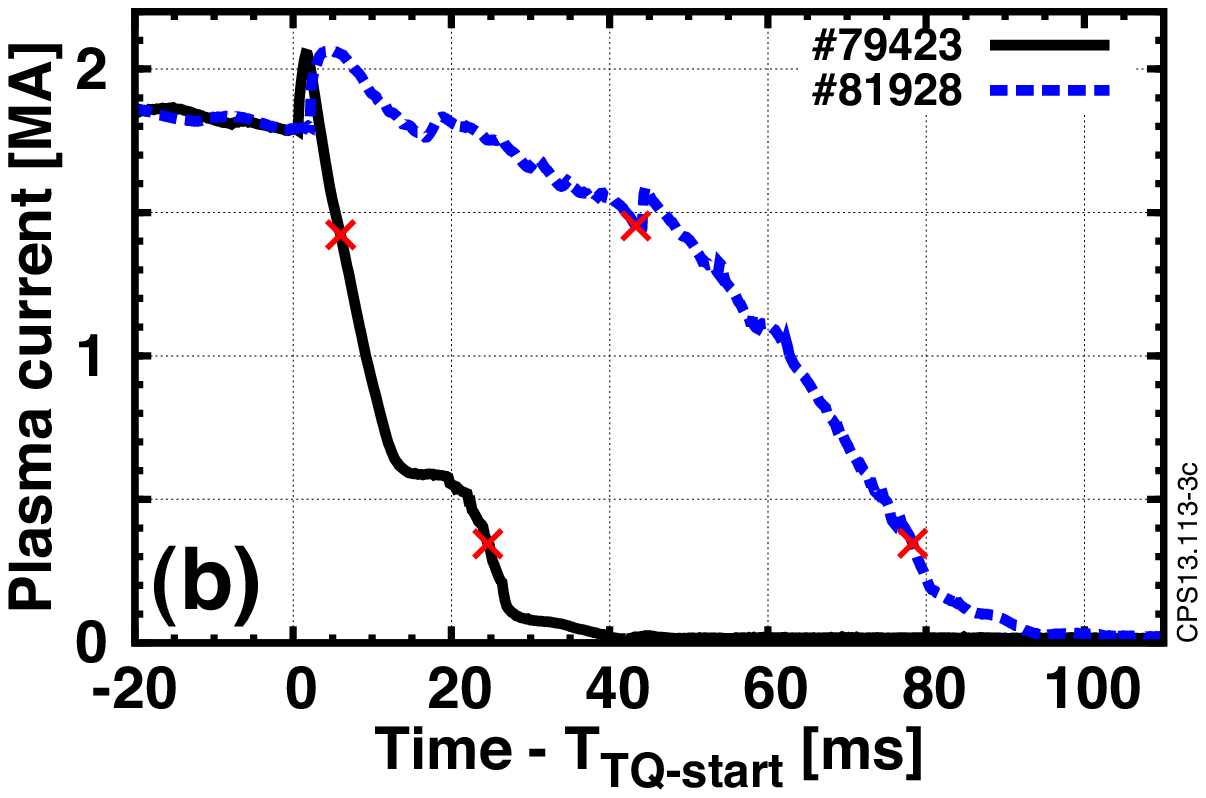}
\caption{\label{fig:T-Ip}{(a) Time evolution of the central electron temperature T$_e$ measured by ECE. (b) Time evolution of plasma current during
    the current quench. Visible plateau for \#79423, slow and steady drop for \#81928. }}
\end{center}
\end{figure}

Figure \ref{fig:T-Ip}a shows the evolution of the central electron temperature. Electron temperatures shown throughout the paper were obtained by the electron cyclotron emission (ECE) diagnostics. Care should be taken in the interpretation of ECE signals as during the disruption the plasma may become optically thin and suprathermal radiation may be present in the spectrum. Also the plasma was moving upwards during the quench in shot \#81928 and therefore the displayed ECE measurement does not exactly represent the core temperature 50 ms after the thermal quench.
For easier comparison, the time axis is now shifted with the thermal quench (TQ) time ($t_{\mathrm{TQ-start}}$ = 22.005~s for \#79423 and $t_{\mathrm{TQ-start}}$ = 20.5357~s for \#81928). The thermal quench occurs at slightly different times with respect to the Ar valve trigger,
although the difference of 30 ms is relatively small compared to the
500 ms delay between the valve trigger and the quench (figure
\ref{fig:inject}).

A closer look in figure \ref{fig:T-Ip}a reveals the main difference in the
otherwise similar temperature collapses. Both disruptions show a relatively mild temperature decrease to
$\sim$1.5 keV in a 40 ms period before the TQ.
In the ILW case the temperature drops down because of increased transport during the thermal quench but then it recovers up to $\sim 300$ eV due to the low radiation intensity and slower transport times during the current quench. The $\sim 300$ eV recovery is followed by a slow drop during a 50 ms timeframe. This very slow drop
results in only a small amount, $\lesssim$ 70 kA of runaways, as was determined by hard X-ray measurements.
In the C wall case the temperature drop after 1.5 keV continues down to $\sim$10 eV and gives rise to runaways with a current plateau of $\sim$600 kA.
Figure \ref{fig:T-Ip}b
shows the evolution of plasma current: a swift current quench with a
runaway plateau at 600 kA in the C wall case, while a slow drop of
current in the ILW case.
The slow drop
in the plasma current is not expected to generate a~sufficiently high
electric field for substantial runaway generation. The differences in the quench times compared to the C wall case are typical with the ILW \cite{devries12impact}.
The observed
disruption- and runaway parameters are given in table
\ref{tab:disruption}. Red crosses in figure \ref{fig:T-Ip}d mark the
80\%/20\% values used to determine the current quench (CQ) time, shown
in table \ref{tab:disruption}.

In the C wall case the steady state \Zeff is 2.2 $\pm$
20\%. Although beryllium discharges in general are considered quite clean, in
discharge \#81928 \Zeff was in the 2.5 $\pm$ 20\% range before the
thermal quench. These discharges were part
of a disruption session with limiter plasmas, where the relatively low plasma density is coupled with increased wall sputtering.
There is a clear exponential rise of the measured \Zeff in both cases during the
thermal quench, but the reliability of \Zeff measurements during
the disruption is low and values during and after the TQ have to be
considered with caution.

\begin{table}
\caption{\label{tab:disruption}The disruption- and runaway parameters in the two discharges.}
\begin{indented}
\item[]\begin{tabular}{@{}llll}
\br
Parameter &\#79423 (CFC)&\#81928 (ILW)\\
\mr
Valve trigger time & 21.51 s & 20.01 s\\
Total pre-disruption & 8.45$\cdot 10^{20}$ & 1.02$\cdot 10^{21}$\\
\hspace{0.5cm}plasma electrons & & \\
\textbf{Total} Ar injected & 7.39$\cdot 10^{20}$& 5.68$\cdot 10^{20}$\\
Ar injected up to TQ & 6$\cdot 10^{20}$& 5.07$\cdot 10^{20}$\\
Ar \% at 100\% assimilation &\textbf{88\%} & \textbf{56.8\%} \\
Thermal quench start & 22.0054 s & 20.5357 s\\
Quench delay & 495.4 ms & 525.7 ms\\
Radiated energy (in 100 ms)&$\sim$5.5 MJ & $\sim$4 MJ\\
CQ time (80\% $\rightarrow$ 20\%) &$\sim$18.6 ms &$\sim$35.0 ms\\
Runaway current & $\sim$600 kA & < 70 kA\\
Runaway plateau & $\sim$6 ms & $\varnothing$ \\
\br
\end{tabular}
\end{indented}
\end{table}

\section{Numerical model}
\label{sec:model}

Runaway electrons in cooling plasmas can be generated by various
mechanisms: Dreicer generation \cite{connor75relativistic}, hot tail
generation \cite{harvey00runaway} and runaway avalanching
\cite{rosenbluth97theory}.
The time evolution of the current density
profile is determined by the runaway electron generation and the
diffusion of the electric field governed by the parallel component of
the induction equation
\begin{equation}
\frac{1}{r}\frac{\partial}{\partial r}\left(r \frac{\partial E}{\partial r}\right)=\mu_0 \frac{\partial}{\partial t}(\sigma_\parallel E+ n_{r} e c),\label{eq:ind}
\end{equation}
where $n_{r}$ is the number density of the runaways -- travelling with
approximately the speed of light -- and $\sigma_\parallel$ is the Spitzer
conductivity with a neoclassical correction \cite{wesson04tokamaks}.
The changes of the (\ref{eq:ind}) electric field are mainly determined by the short time
scale changes of the conductivity, which strongly depends on temperature ($\sigma \propto T^{3/2}$).
The model also includes a conducting plasma vessel \cite{smith06go,feher11simulation} but neglects coupling to the coils.
In the thermal quench the
conductivity drops and that induces a rising electric field which
gives rise to a seed population via the Dreicer process
\begin{equation} 
  \left(\frac{\ud n_{r}}{\ud t}\right)_{D} \simeq
  \frac{n_{e}}{\tau} \left( \frac{m_{e} c^2}{2
      T_{e}} \right)^{3/2} \left( \frac{E_{D}}{E}
  \right)^{3(1+Z_{\rm eff})/16}
  e^{-\frac{E_{D}}{4E}-\sqrt{\frac{(1+Z_{\rm eff})E_{D}}{E}}}.\label{n1D}
\end{equation}
Here, $E_{D} = m_{e}^2 c^3 / (e \tau T_{e})$ is the Dreicer field, and
$\tau$ is the relativistic electron collision time $\tau = 4 \pi
\varepsilon_0^2 m_{e}^2 c^3 / (n_{e} e^4 \ln \Lambda)$ and
$\ln\Lambda$ is the Coulomb logarithm. The seed runaways are
amplified via avalanching \cite{rosenbluth97theory}:
\begin{eqnarray}
  \left(\frac{dn_{r}}{dt}\right)_{\mathrm{avalanche}} &\simeq & n_{r}
  \frac{E/E_{c}-1}{\tau \ln \Lambda}\sqrt{\frac{\pi \varphi}{3(Z_{\rm
        eff}+5)}}\times \nonumber\\ && \left( 1-
    \frac{E_{c}}{E}+\frac{4\pi (Z_{\rm eff}+1)^2}{3\varphi (Z_{\rm
        eff}+5)(E^2/E_{c}^2+4/\varphi^2-1)} \right)^{-1/2},
  \label{2_generation}
\end{eqnarray}
where $E_{c}=m_{e}c/(e\tau)$  is the critical electric field, $\varphi=(1+1.46\epsilon^{1/2}+1.72 \epsilon)^{-1}$ and
$\epsilon=r/R$ denotes the inverse aspect ratio.
There are several processes that can limit the energy of
the runaways or contribute to their losses, such as radial diffusion
due to magnetic perturbations, synchrotron radiation
\cite{andersson01damping}, bremsstrahlung, or plasma instabilities
driven by the runaway beam anisotropy \cite{fulop09magnetic}. In this work, we
consider losses due to radial diffusion using the Rechester-Rosenbluth diffusion estimate
\cite{rechester78electron} $ D_{RR} =
\pi q v_\| R \left(\delta B/B\right)^2 $, where $v_\parallel \simeq c$ is the parallel velocity, $R$ is the major
radius and $\delta B/B$ is the normalized magnetic perturbation
amplitude. This numerical tool (called the GO code) was initially presented in
\cite{eriksson04current,smith06go} and developed further in
refs.~\cite{feher11simulation,gal08runaway}. See reference \cite{feher11simulation} for further details and parameter scans.
In the version of the GO code used in this paper, the Dreicer and
avalanche runaway rates and radial losses due to magnetic field
perturbations are coupled to the evolution of the electric field
through equation (\ref{eq:ind}).
Hot tail generation is efficient if the
cooling rate is comparable to the collision frequency
\cite{helander04electron} and has been predicted to be important in ITER
disruptions \cite{smith05runaway}, but in the cases studied in this
work, the cooling times are long enough for the Dreicer generation to
dominate over hot-tail generation.

The GO code requires specification of the neutral impurity density as function of
time and radius, $n_{Z_i}^0(r,t)$. The time evolution is often assumed to be an exponential ramp-up, with a characteristic time on the ms
timescale in agreement with numerical modelling \cite{izzo11runaway}.
The temperature and density evolution is modeled separately for each plasma component -- electrons and $Z_i$ ions.
The energy balance equations describing the temperature evolution are
\begin{eqnarray}
    \frac{3}{2}\frac{\partial  (n_{e} T_{e})}{\partial t}&=\frac{3
      n_{e}}{2 r}\frac{\partial
    }{\partial r}\left(\chi r \frac{\partial T_{e}}{\partial
        r}\right)+P_{\rm OH}-P_{\rm line}-P_{\rm Br}-P_{\rm ion}+\sum_i P_{\rm c}^{eZ_i}, \\
    \frac{3}{2} \frac{\partial  (n_{Z_i} T_{Z_i})}{\partial
      t}&=\frac{3n_{Z_i}}{2 r}\frac{\partial
    }{\partial r}\left(\chi r \frac{\partial T_{Z_i}}{\partial
        r}\right)+P_{\rm c}^{Z_ie}+\sum_{j\neq i}P_{c}^{Z_iZ_j}.\label{eq:ebalance}
\end{eqnarray}
Here $P_{\rm OH}=\sigma_\|E^2$ is the Ohmic heating power density, $P_{\rm line}$ and 
$P_{\rm Br}$ are the line- and Bremsstrahlung radiation and $P_{ion}$
is the ionization energy loss. Bremsstrahlung losses are taken into account with the formula $P_{\rm Br}= 1.69\cdot 10^{-38} n_{e}^2
\sqrt{T} Z_{\rm eff}$ \cite{NRL}.
Due to the different collision times the different species are modeled separately.
The (\ref{eq:ebalance}) energy balance equations are coupled with collisional
energy exchange terms between Maxwellian species
\cite{helander02collisional}: $P_{{\rm c}}^{ij}=3n_i(T_j-T_i)/2\tau_{ij}$
with the heat exchange time
$$\tau_{ij}=\frac{3\sqrt{2}\pi^{3/2}\epsilon_0^2 m_i m_j}{n_j e^4 Z_i^2 Z_j^2 \ln{\Lambda}}
\left(\frac{T_i}{m_i}+\frac{T_j}{m_j}\right)^{3/2},$$ where the subscripts
$i,j$ now refer to electrons as well as deuterium \& impurity ions.
The heat diffusion coefficient is assumed to be constant ($\chi=1$ m$^2/$s) unless otherwise indicated. Studies were made in ref.~\cite{feher11simulation} to test
the influence of this assumption on the GO simulation results.
Radiation has the strongest cooling effect on the electrons.
To describe the line radiation we calculate the ionization of the impurities by
calculating the density of each charge state for every ion species ($n^k_{Z_i}, k=0..Z_i$):
\begin{equation*}
    \frac{\ud n_{Z_i}^k}{\ud t}=  n_{e} \left( I_{k-1} n_{Z_i}^{k-1} - (I_k+ R_k) n_{Z_i}^{k} +  R_{k+1} n_{Z_i}^{k+1} \right),
\end{equation*}
where $I_{k}$ denotes the electron impact ionization rate for the
$k$-th charge state and $R_k$ is the radiative recombination rate
\cite{NRL}.  The line radiation is calculated by
\begin{equation*}
  P_{{\rm line}}=\sum_i n_{Z_i} n_{e} L_{Z_i}(n_{e},T_{e}).
\end{equation*}
The radiation rates $L_{Z_i}(n_{e},T_{e})$ are extracted from the ADAS database \cite{smith04adas}. We note that from a numerical point of view this ionisation / recombination \& radiation calculation is the most CPU intensive task as each transition for every charge state in every ion species has to be calculated in every time step.

\section{Results and discussion}
\label{sec:results}

The initial plasma parameters used in the simulations are given in
table \ref{tab:parameters}. Apart from these, the simulations have a
number of input parameters that cannot be solidly based on the
experimental data, mostly due to the fact that several quantities are
extremely hard to accurately determine during a disruption. Knowledge
about the impurity content of the plasma, the mixing efficiency and
the impurity mixing time is lacking and therefore comparison of the
simulated spatio-temporal distribution of impurities is not
possible. There are, however, integrated measurements available
with reasonable accuracy to account for the impurity penetration,
namely the evolution of plasma temperature, density, \Zeff or radiation. Our
simulations using predescribed plasma parameter evolution show that change in plasma electron density with the
experimentally measured magnitude of $\sim 30$\% has a negligible effect on the saturation runaway current.
Plasma temperature, especially its temporal
evolution is more important.
As will be demonstrated in this section, also \Zeff has a significant effect on runaway
generation. \Zeff measurements have $\sim$20\%
uncertainty before the disruption, and after the disruption the
reliability is not good. In the following we will start with
investigating the effect of the temperature evolution (setting
\Zeff=1) and in the next subsection we will model the impurity
injection, including scans of argon and background impurity (carbon or
beryllium) content to investigate the effect of radiation and \Zeffd

\subsection{Predescribed temperature evolution}

To determine how well the experimental measurements can be connected
to the complex simulation of self-consistent current-, runaway
electron-, electric field- and impurity evolution, a somewhat simpler
approach is used at first. 
The evolution of plasma temperature, density and \Zeff is prescribed and impurity evolution is not followed. This allows us to understand how these parameters affect the runaway evolution before turning to more sophisticated simulations where the evolution of the impurities determine the aforementioned quantities.
$T_e(t)$ is obtained by either fitting or
directly using the experimental data (with interpolation) as an
input. The latter seems to be simpler and more realistic but we have
to address $T_e < 0$ experimental values and the possible ambiguity of ECE $T_e$ data. To compensate for $T_e < 0$ points and the upwards plasma movement in shot \#81928 the data used 50 ms after the thermal quench is elevated based on the moving average of the experimental scatter of the recorded temperature. This correction is in the order of 10 eV in the plasma center 50 ms after the disruption and decreases gradually as the temperature decreases.
In discharge \#79423 the final electron temperature
is in the order of 10 eV and the characteristic drop time is
$\tau_0=0.26$ ms.
Before trying to answer {\em why} the temperature evolution
looks as it does for the two cases we wish to understand the
implications of such temperature evolution on runaway electron
generation. 

\begin{figure}[htb!]
\begin{center}
\includegraphics[height=51mm]{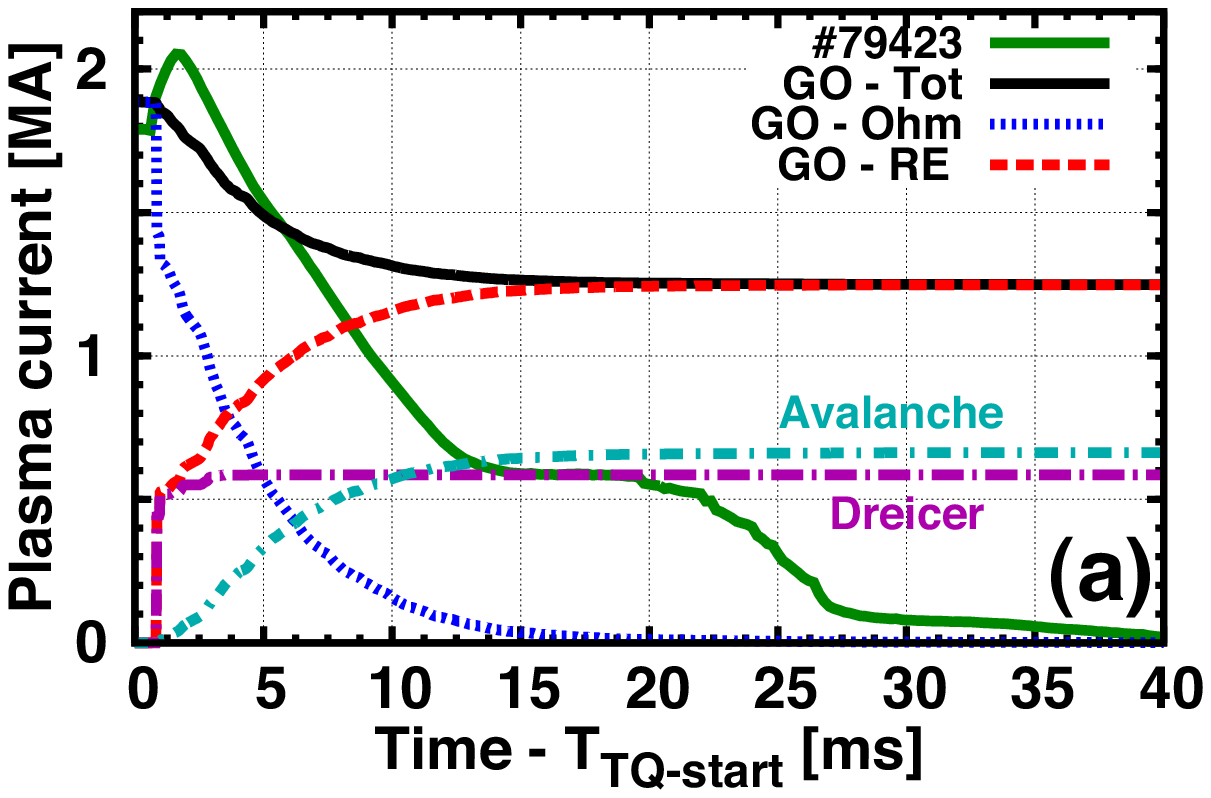}\hfill
\includegraphics[height=51mm]{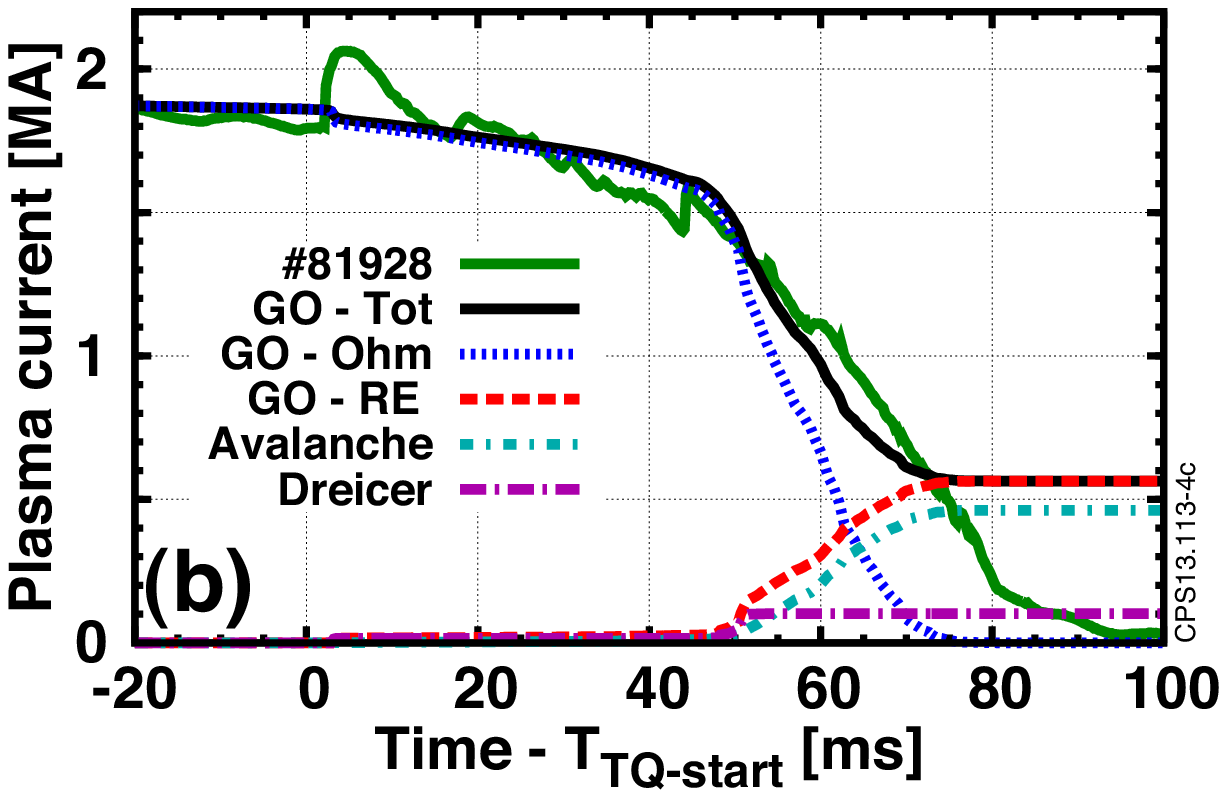}
\caption{\label{fig:tdata79423-81928}{Evolution of different current
    components vs the measured plasma current evolution for (a) discharge
    \#79423, (b) \#81928. $T_e(t)$ taken from the ECE data.}}
\end{center}
\end{figure}

First we consider the C wall case. Figure \ref{fig:tdata79423-81928}a shows
the evolution of different current components for \#79423 together with the
experimental current evolution (shown with green solid line) for the
case when $T_e(t)$ is taken from the ECE data.
In these simulations, for simplicity \Zeff is set to 1. With increasing \Zeff the runaway current increases as
will be shown later. The simulated current evolution is largely different from the experimental one. The runaway
current in the discharge was $\simeq$590 kA while in the simulated
case it is 1.27 MA. One of the reasons for these differences is
that at this point the evolution of plasma density, temperature and impurity profiles are not followed self-consistently.
Another reason is that in the case of figure \ref{fig:tdata79423-81928}a-b
no runaway losses are included in the numerical
calculation, which would certainly be present during e.g. a violent MHD mixing
scenario \cite{izzo11runaway}, error fields from the coils or movement towards the wall. The effect
of losses due to magnetic perturbations will be assessed later in section \ref{subsec:diff}. The Dreicer current represents $\sim$47\% of the total RE current.

\begin{figure}[htb!]
\begin{center}
\includegraphics[height=51mm]{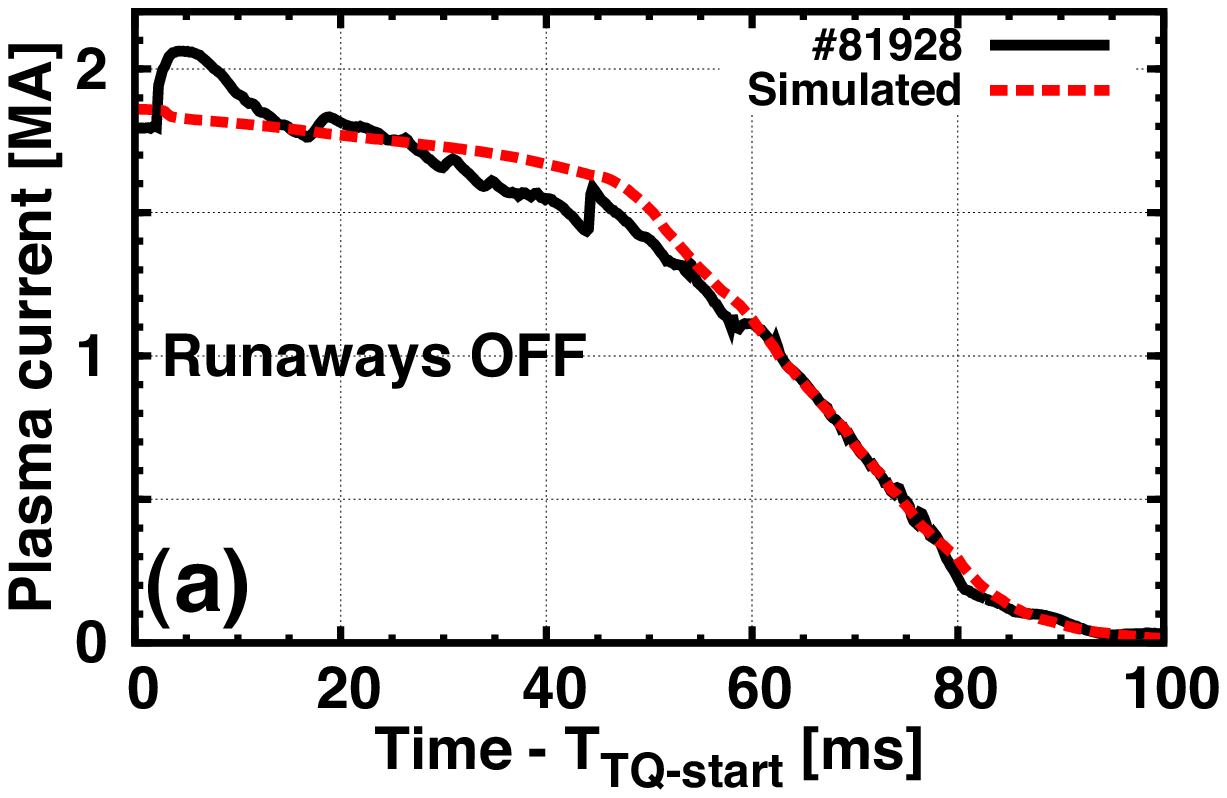}\hfill
\includegraphics[height=51mm]{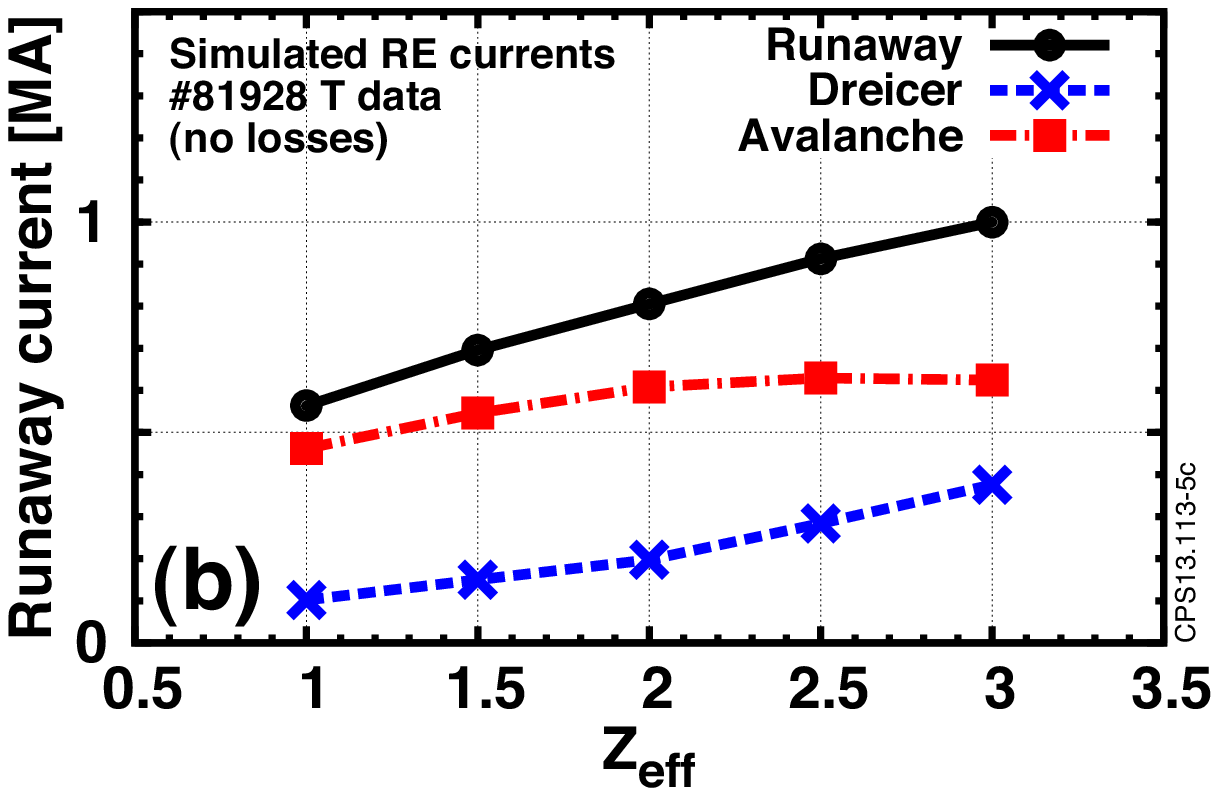}\hfill
\caption{\label{fig:tfit81928norun}{(a)\#81928. Good agreement of simulated and experimental current evolution if runaway generation
    is switched off. (b) Runaway current as a function of a uniform, constant \Zeff in the \#81928 case.}}
\end{center}
\end{figure}

For the ILW case, which is shown in figure \ref{fig:tdata79423-81928}b,
the usage of the corrected experimental temperature data leads to the generation of runaway current in the order of 4-500 kA, the majority of which generated by avalanching. This can be overcome by runaway losses as will be shown in section \ref{subsec:diff}.
If we switch off the runaway generation in the
ILW case \#81928 and follow the current decay due to the
temperature drop including the aforementioned correction, we find a very good agreement between the
simulated and the measured plasma current evolution (figure
\ref{fig:tfit81928norun}a). We have to note that in shot \#81928 the central temperature measurement after 50 ms is highly uncertain, but with the application of the aformenetioned data correction the simulated current evolution matches the experiment.

The runaway current obtained in the
simulations excluding losses is too large, even in the case of \Zeff=1.
Increasing \Zeff will lead to even larger runaway currents. Figure
\ref{fig:tfit81928norun}b shows the runaway current as a function of a
uniform, constant \Zeff with the temperature evolution of discharge
\#81928. The runaway current is rapidly increasing with
\Zeffd~ Therefore, it is clear that more sophisticated modelling,
including the effect of the injected impurities and losses are required to
understand the experimental result of suppressed runaway current in the case of
ILW.

\subsection{Impurity injection}

\begin{figure}[htb!]
\begin{center}
\includegraphics[width=0.49\linewidth]{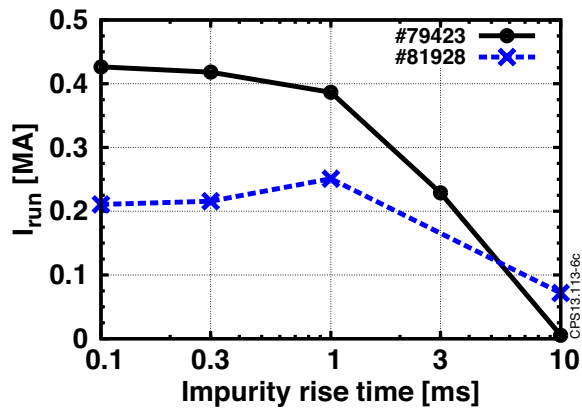}
\caption{\label{fig:scan3_t0}{Runaway current for different impurity rise times with 20\% Ar injected.}}
\end{center}
\end{figure}

In this section the GO code is used in a full mode, where the evolution of plasma- and impurity parameters are calculated in a self-consistent simulation.
To model the impurity radiation and ionization \& recombination processes and their effect on temperature, \Zeff and density; the GO code
requires specification of the neutral impurity density as function of
time and radius. As the impurities penetrate the plasma, the energy balance equation is solved taking into account radiation, ionisation, recombination, collisions and heat diffusion \cite{feher11simulation}. This calculation determines the temperature, \Zeff and density profiles, which influence the evolution of the plasma- and runaway current profiles. The evolution of the currents act back on the plasma- and impurity dynamics through Ohmic heating.
In this paper, the impurity density radial profile
shape is assumed to be equal to the pre-disruption electron density
profile shape and its absolute value is ramped up exponentially in time to its
final value with a~predescribed characteristic rise time.  The
impurity contents are defined as the ratio of the total number of
injected neutral impurity atoms to the initial total electron content (see table \ref{tab:disruption}).
For most of the cases we used 0.3 ms as the exponential time constant,
which puts the characteristic time of the impurity evolution on the ms
timescale in agreement with numerical modelling
\cite{izzo11runaway}. With longer impurity mixing time the quench is
less energetic as the various heating mechanisms can better counteract
the energy loss due to ionisation and radiation. This in turn leads to
a drop in the generated runaway current, as is shown in figure
\ref{fig:scan3_t0}.

\begin{figure}[htb!]
\begin{center}
\includegraphics[width=0.49\linewidth]{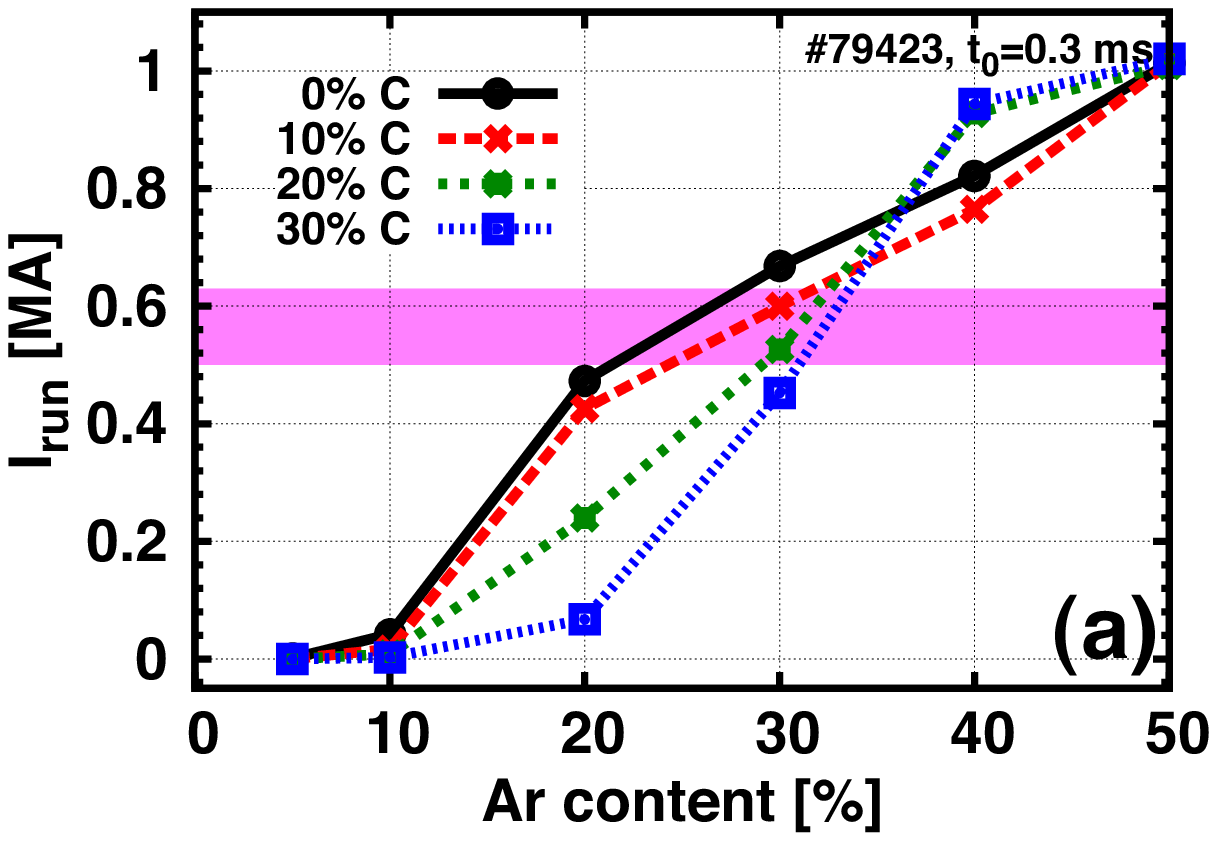}\hfill
\includegraphics[width=0.49\linewidth]{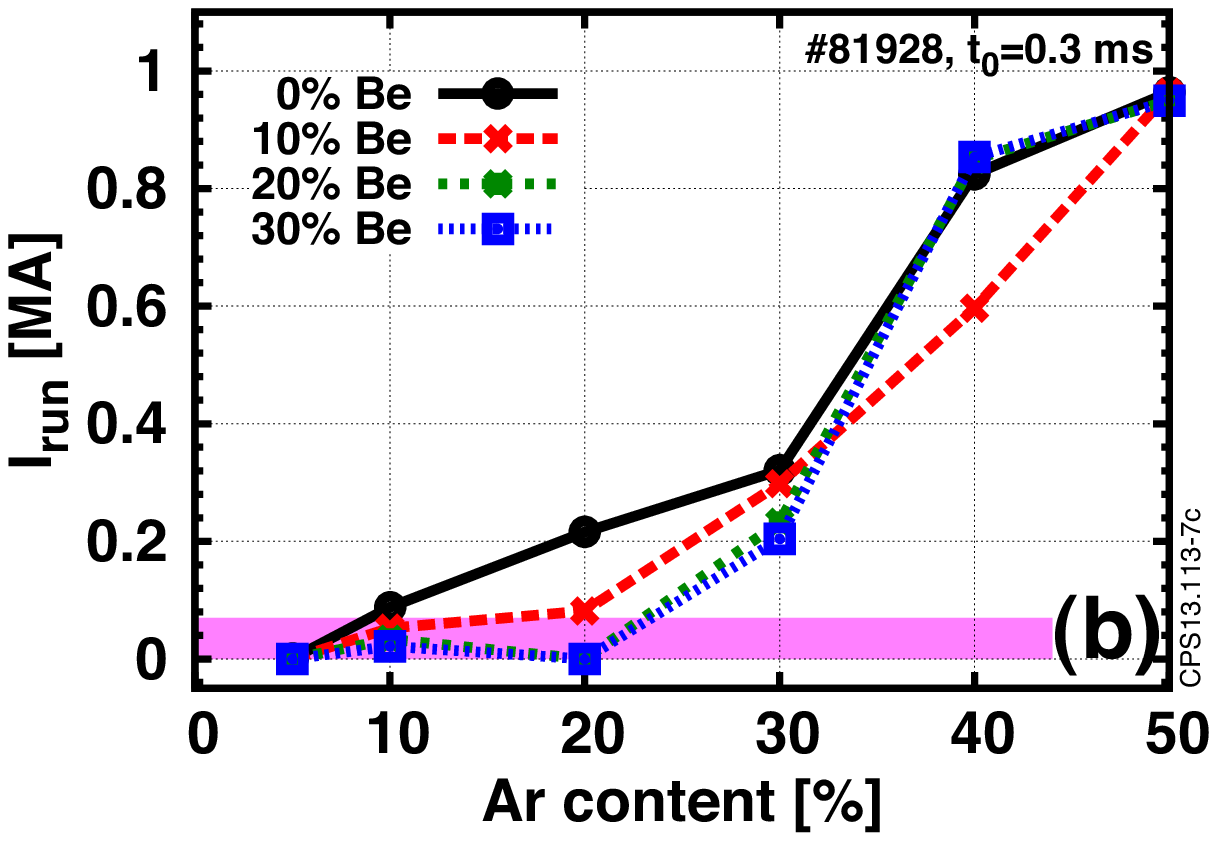}
\caption{\label{fig:scan1_Irun}{Runaway current for various Ar and (a)
    C or (b) Be contents in the representative discharges. The magenta
    rectangle marks the experimentally measured range of runaway
    current.}}
\end{center}
\end{figure}

We have carried out scans for the argon and background impurity
content to assess the similarities or differences in the runaway
behaviour in the two model discharges.  The argon content can be
estimated based on the total injected neutral argon amount (see table
\ref{tab:disruption}) with a
reasonable assumption for the mixing efficiency, that is expected in the order of 30\% \cite{hollmann08measurement, reux13private}.
The relative amount of total argon injected in
the C wall case discharge \#79423 was roughly 30\% more than that of
discharge \#81928. If we assume a similar mixing efficiency in the two
cases, then the difference in the injected amount should reflect in a
larger argon content of the plasma during the disruption in the C wall
case. Pre-disruption Be/C levels can be estimated from the pre-disruption \Zeff value
(table \ref{tab:parameters}). However, the level of impurity
sputtering during the disruption is unknown and therefore we scanned up to 30\% wall impurity content.

Figure \ref{fig:scan1_Irun}a shows the effect of argon and carbon
content on the obtained runaway currents. As a general trend the
runaway current increases with argon content, while it decreases with
carbon content. The magnitude of the latter effect depends also on the
argon content. The magenta rectangle represents the experimentally
measured runaway current, which puts the argon content between 20-30\%
and the carbon content in the range of 10-30\%.
Note that 7\% of fully ionized carbon would lead to \Zeff~$\simeq$~2.5.
20-30\% argon content is reasonable considering the injected argon amount and $\sim$30\% mixing.
Figure \ref{fig:scan1_Irun}b shows the effect of argon and
beryllium. The main trend of increasing runaway current with argon
content is basically the same, but the exact numbers are
different. This is due to the nonlinear nature of the simulations that
amplify the differences in the initial temperature- and density
profiles as well as due to the presence of different background
impurities. The presence of beryllium effectively reduces the runaway
current at argon contents of experimental relevance (<20\%). As low as
10\% beryllium (corresponding to \Zeff = 1.65) leads to a factor of
two decrease in runaway current. The experimentally measured $\lesssim 70$ kA
runaway current is therefore reproduced at reasonable impurity
contents. Comparing the 0\% wall impurity cases for the same argon amounts in figures \ref{fig:scan1_Irun}a-b reveals the sensitivity of runaway generation to the differences in the initial parameters. Shot \#81928 shows a large reduction in the runaway current for argon contents between 5\%--40\%. Above 40\% the runaway behaviour is similar in the two shots.
This shows that the experimentally observed differences with the ILW are in part caused by the differences in plasma parameters. This effect is further enhanced by the different injected argon amount and the effect of wall impurities.
The simulations indicate that the runaway current and Dreicer fraction reducing effect of the wall impurities decreases with increasing Ar content (for the plasma parameters in these shots) and the behaviour is comparable above 50\% argon content. This suggests that runaway electrons may return in future experiments regardless of the ILW when argon is used in large quantities in massive gas injection (MGI) experiments on JET and ITER.

\begin{figure}[htb!]
\begin{center}
\includegraphics[width=0.49\linewidth]{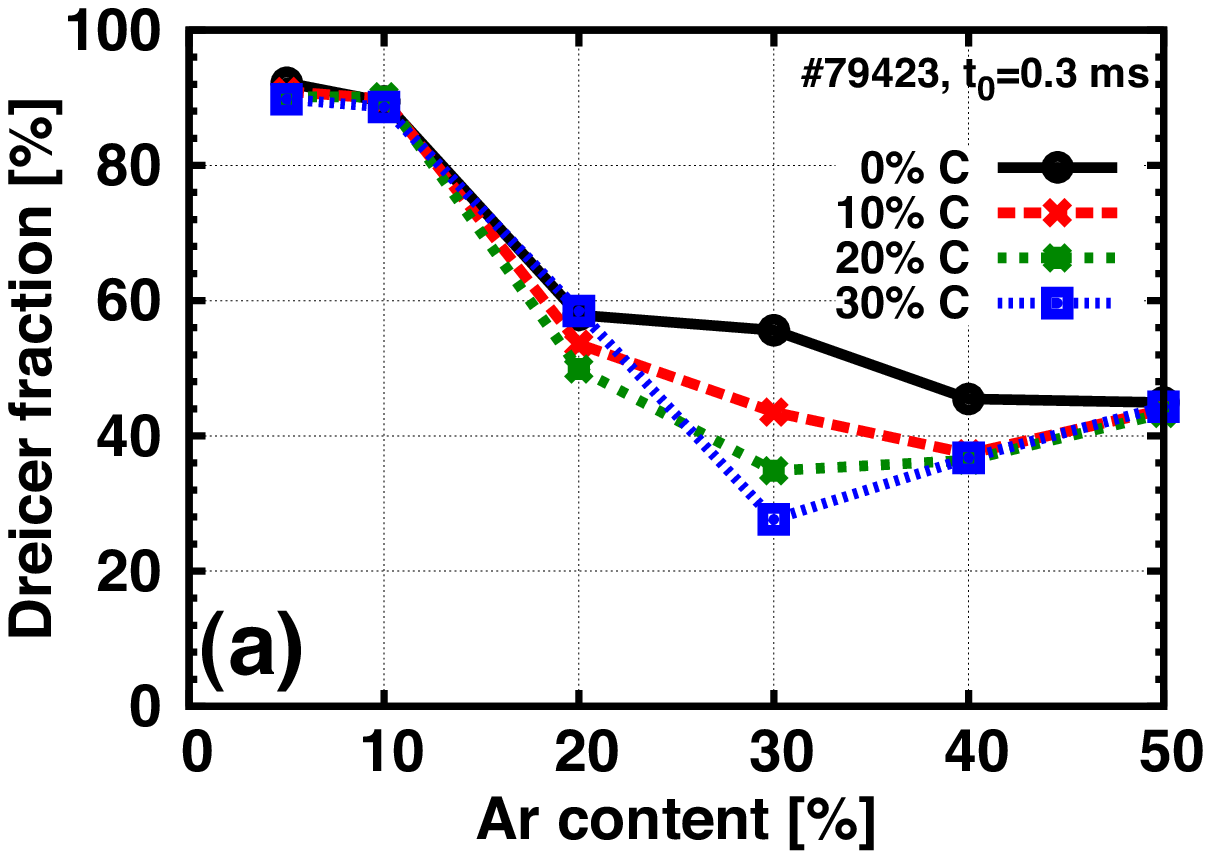}\hfill
\includegraphics[width=0.49\linewidth]{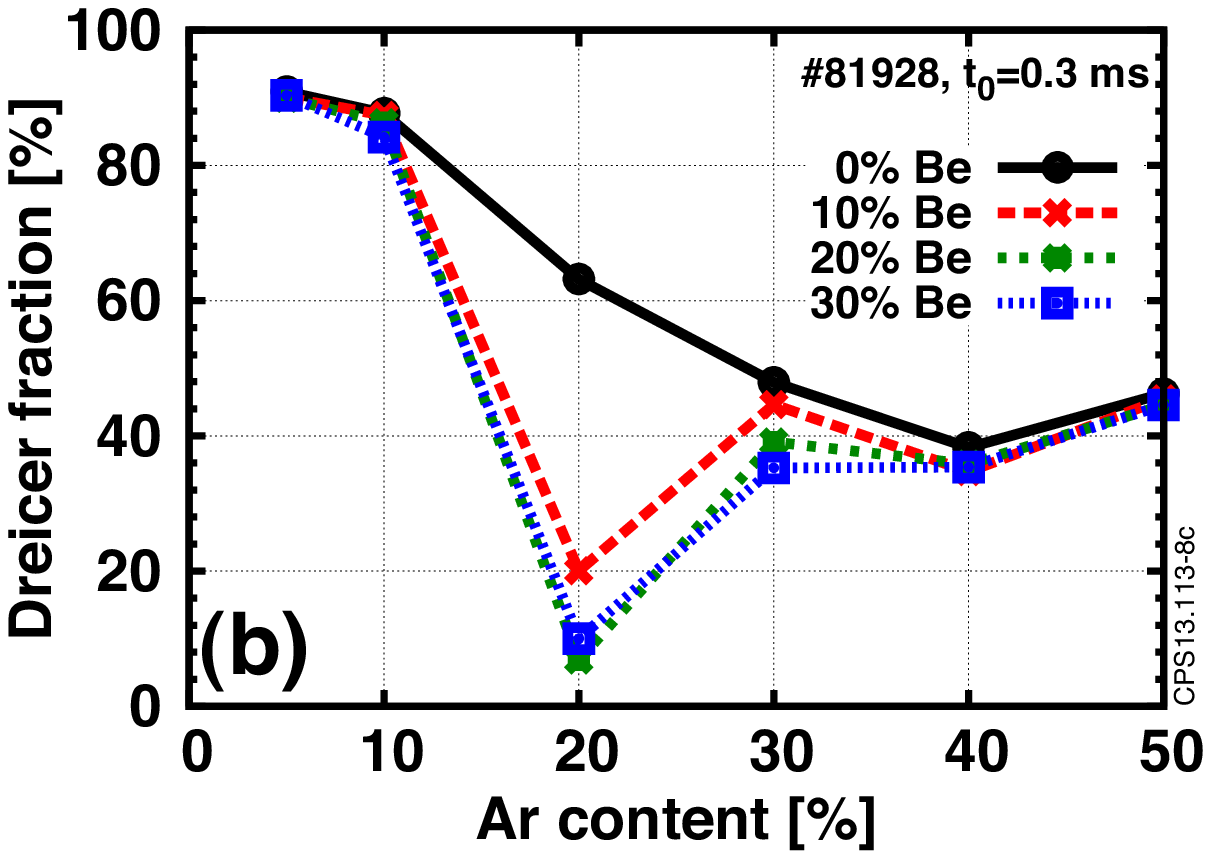}
\caption{\label{fig:scan1_Dfrac}{Dreicer current fraction for various
    argon and (a) carbon or (b) beryllium contents in the
    representative discharges. Relatively small amounts of beryllium
    reduce the Dreicer current fraction drastically.}}
\end{center}
\end{figure}

The strength of various runaway generation mechanisms also depend on
the impurity contents. In figure \ref{fig:scan1_Dfrac}a-b the two main
components (Dreicer- and avalanching) are compared. As a general trend
avalanching is more pronounced with increasing argon
quantities. Carbon only modifies the Dreicer current fraction at
$\sim$30\% argon content. The change in the beryllium content causes a more than $40$\% drop
in the Dreicer contribution (and in the total runaway current) at
around 20\% argon content. As the growth of avalanche RE current is slower than Dreicer RE current, cases with low Dreicer fraction are more sensitive to runaway losses. This will be discussed later in section \ref{subsec:diff}.

\begin{figure}[htb!]
\begin{center}
\includegraphics[width=0.49\linewidth]{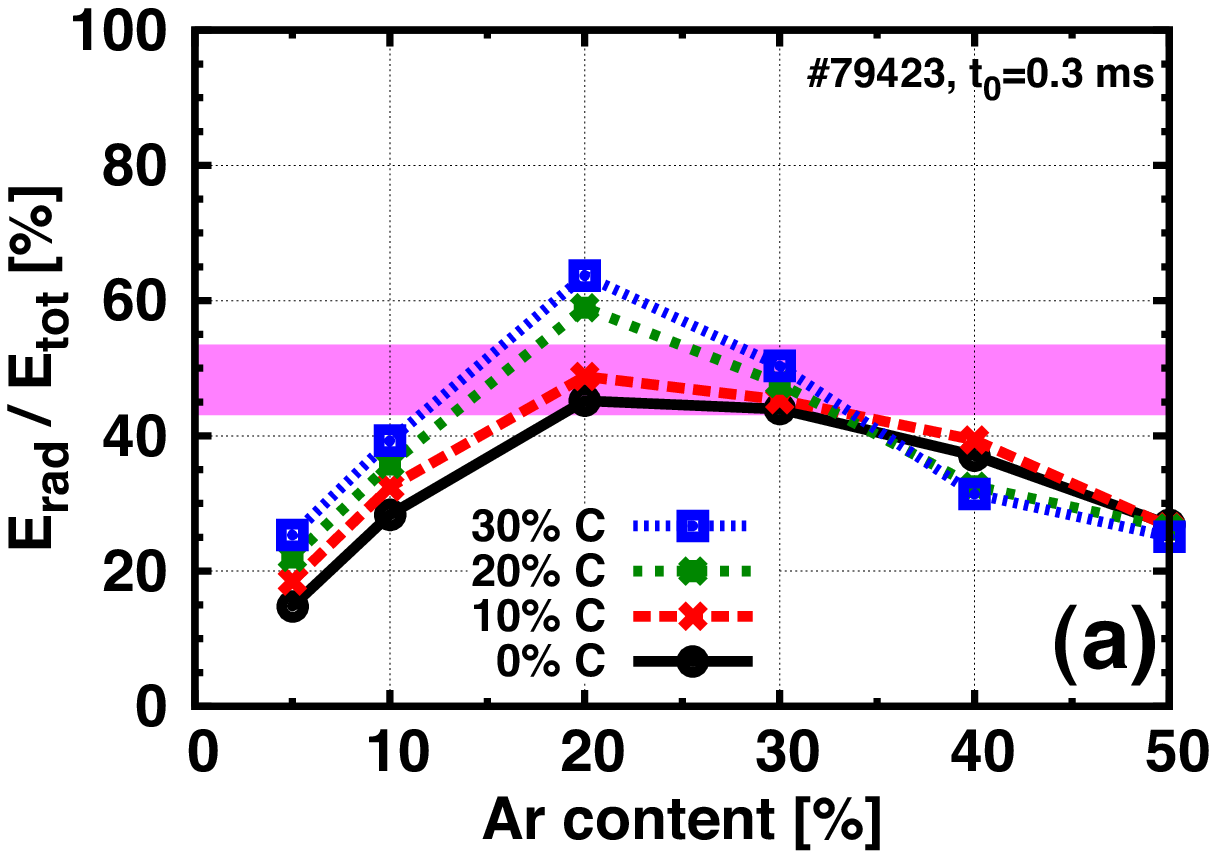}\hfill
\includegraphics[width=0.49\linewidth]{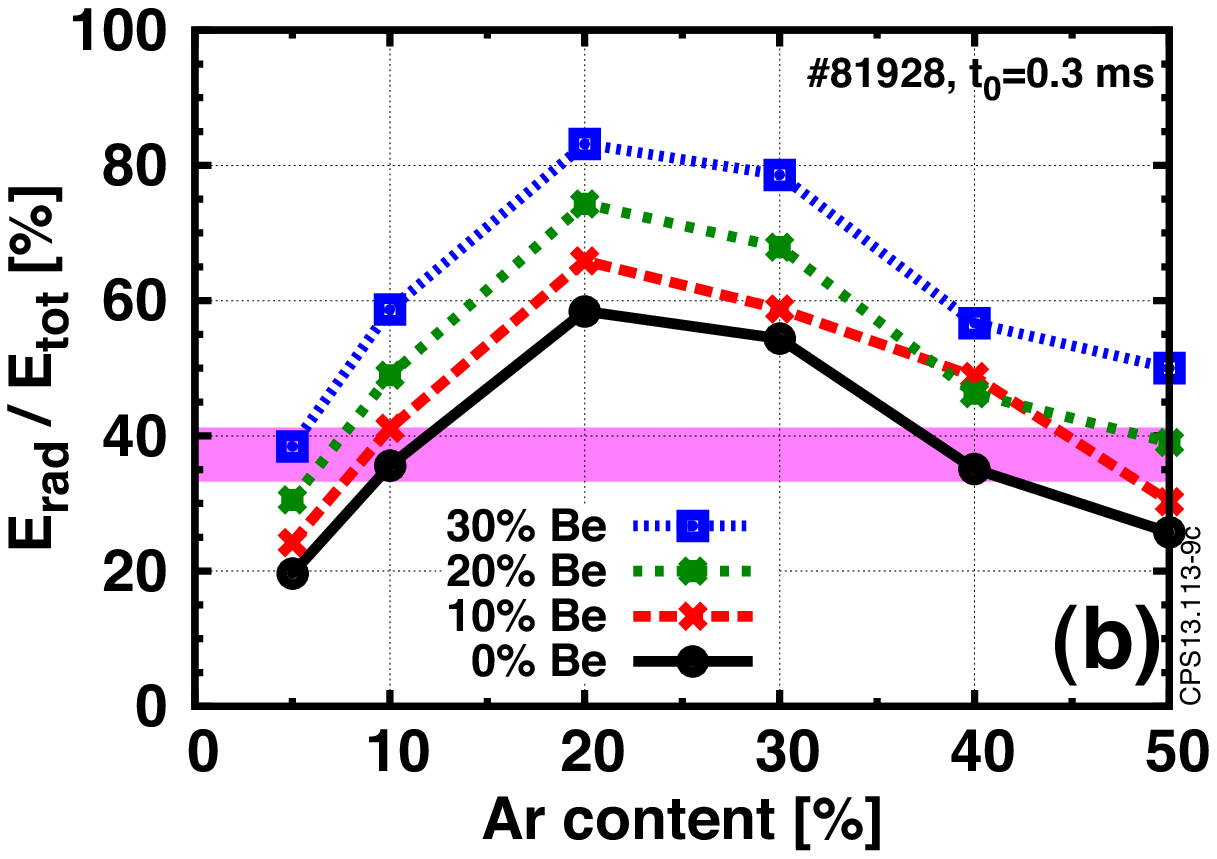}
\caption{\label{fig:scan1_Prad}{Total radiated energy  during a 100 ms
    period after the thermal quench start normalized to the total plasma energy content for various argon and (a)
    carbon or (b) beryllium contents in the representative
    discharges. The magenta rectangle marks the experimentally
    measured total radiated energy fraction for the same period.}}
\end{center}
\end{figure}

The radiated energy during the disruption is estimated using bolometry
measurements. The radiated energy within 100 ms after the thermal
quench is marked with magenta rectangles in figures
\ref{fig:scan1_Prad}a-b. The carbon simulations are in good agreement
with the experiment in terms of radiated energy for argon content
15-30\% with 10-30\% carbon content. The increase in radiated energy
with more carbon is not overly pronounced.  The beryllium simulations
show a steady, up to 50\% increase in radiated energy at higher beryllium
contents. In the beryllium case the \textit{peak} radiation power is much lower
than in the carbon case, but a lower radiation level after the quench
is kept until the plasma current completely decays and this sums up to
a comparable order of magnitude in terms of total radiated energy in these shots.
Experimental agreement is found at different Ar levels for the two shots, a lower amount of Ar is required for the ILW case.
Besides the extra 30\% injected argon in the carbon case of \#79423, the impurity mixing can also be affected by the differences between the two shots.
We have to note that comparison of the simulated and the measured plasma radiation has to be considered with caution.
The total stored energy before the thermal quench is for both pulses  $\sim$11.4 MJ, 10.6 MJ of which is magnetic energy. Depending on the current decay, 42\% (\#79423) / 38\% (\#81928) of the magnetic energy is dissipated in the coils and structure. The maximum amount that could be radiated is therefore 58\% for \#79423 and 62\% for \#81928. The remaining energy which is not radiated or dissipated in the structure is lost by transport to the first wall. The model includes a conducting plasma structure, but not the coils, which should be implemented in future calculations along with a self-consistent handling of plasma movement towards the wall. This is expected to reduce the radiated energy and bring the simulation points closer to the measurement for the ILW case.

\subsection{Diffusion losses}
\label{subsec:diff}

\begin{figure}[htb!]
\begin{center}
\includegraphics[height=51mm]{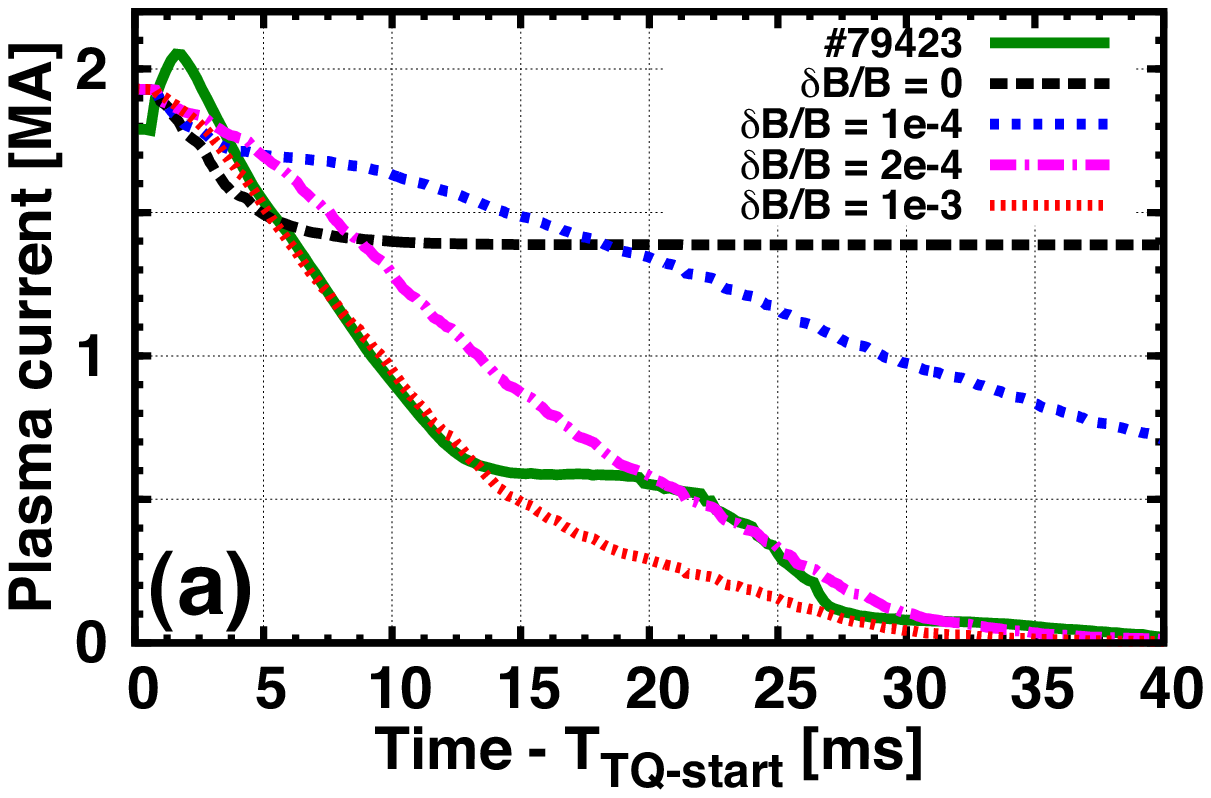}\hfill
\includegraphics[height=51mm]{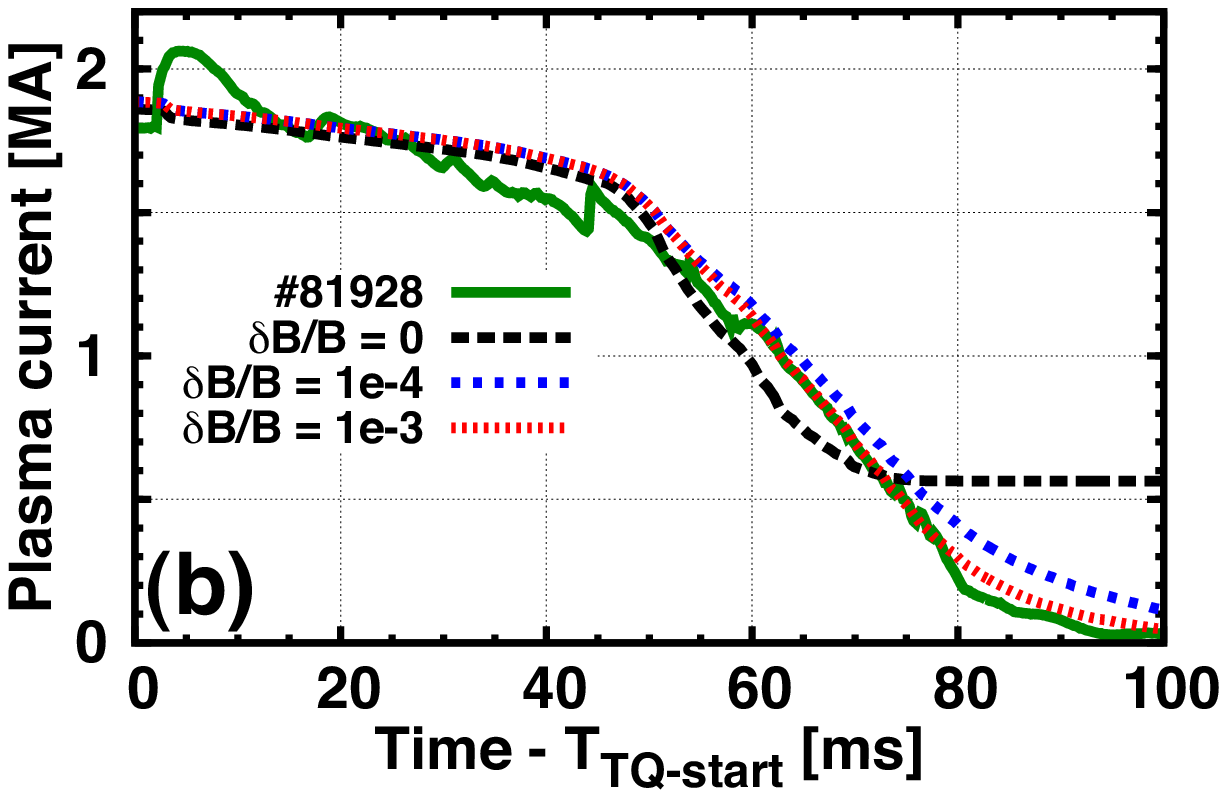}\\
\includegraphics[height=51mm]{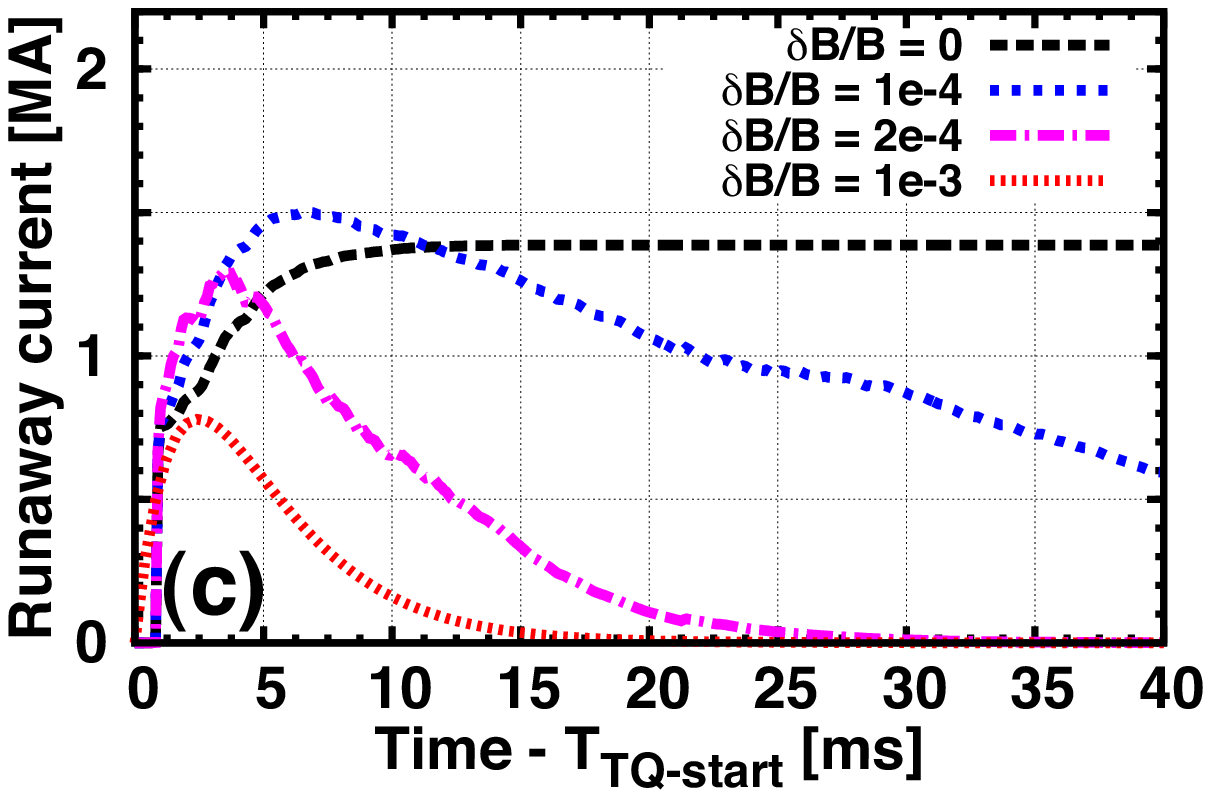}\hfill
\includegraphics[height=51mm]{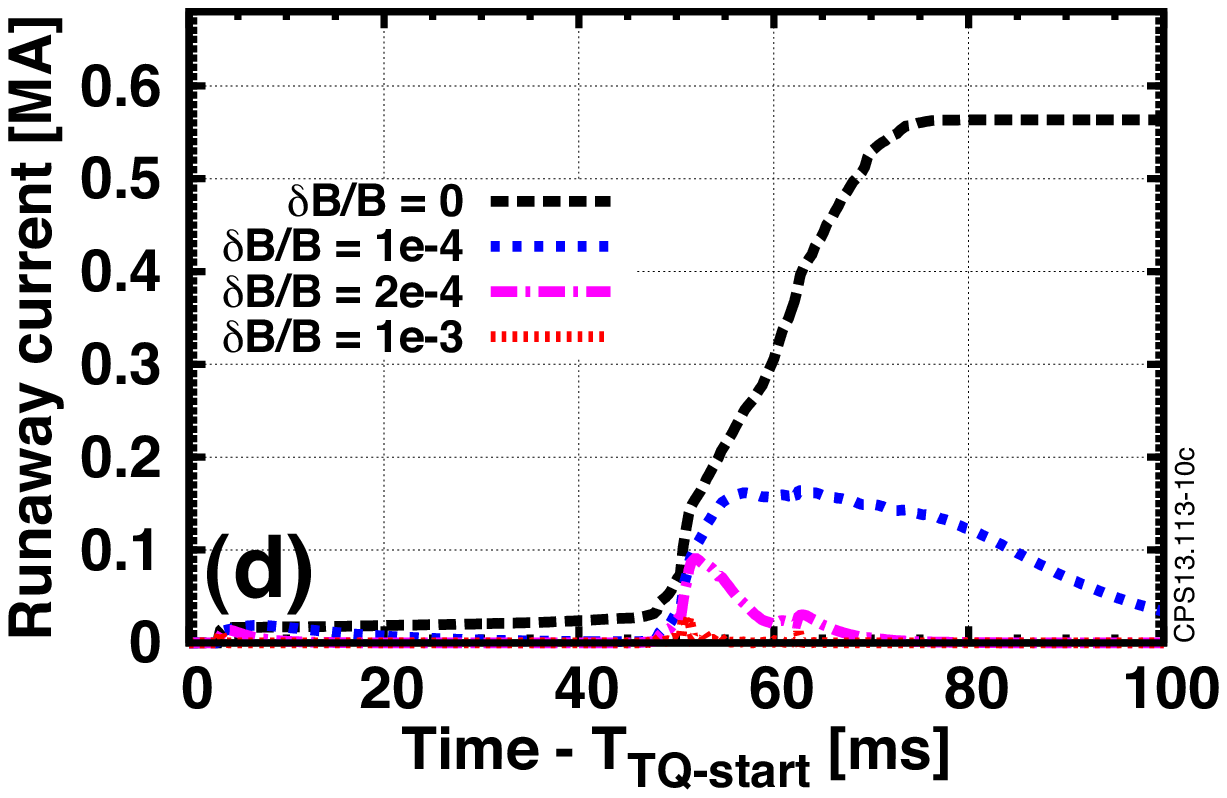}
\caption{\label{fig:dB1}{The effect of magnetic perturbations on plasma current evolution during disruption simulations. (a) C wall case (b) ILW case. Solid (green) lines show the measured plasma current. (c)-(d) RE current evolution corresponding to (a)-(b).}}
\end{center}
\end{figure}

Magnetic perturbations can reduce the runaway electron density, as it
has been shown in theoretical and numerical studies
\cite{feher11simulation,izzo11runaway,helander00suppression,papp11runaway,papp11iter,papp12iter}. Magnetic perturbations can come from the MHD mixing, error fields, instabilities enhanced by the current gradients during runaway evolution, etc. However, accurate measurement of magnetic perturbations during a disruption, especially core perturbations, is extremely challenging.
In this section we will demonstrate the effect of magnetic perturbations on runaway current evolution for the two shots investigated in the paper.
Figure \ref{fig:dB1}a-b shows the effect of magnetic perturbations on the
plasma current in the two cases, for various values of $\delta
B/B$. As we are only interested in the effect of the magnetic
perturbation, in these simulations, we take the temperature from the
experiment, as in subsection 4.1 (and do not simulate the impurity
injection itself). Without runaway losses due to magnetic
perturbations, the simulation ends with a considerable runaway current
in both cases, although it is higher in the C wall case (figure
\ref{fig:dB1}a) than in the ILW case (figure \ref{fig:dB1}b). When the
magnetic perturbation level is increased up to $10^{-4}$, radial
diffusion makes the runaway beam broader but some
of the runaway current still persists. The current
evolution is best matched with a perturbation level between $\delta
B/B= [0.2$ -- $1]\times 10^{-3}$.
 Note, that with constant magnetic perturbation level in the
simulations, we do not expect that the experimentally observed current
plateau should be reproduced. In reality the magnetic perturbation
level depends on time and space and therefore these simulations serve
only to show the magnitude of the effect. 
With a perturbation level of
$\delta B/B=10^{-3}$ the runaway loss rate is comparable to the
generation rate and the runaways spread out in the plasma before they
can form a strong runaway beam.  Even if the runaways are not
completely removed, the runaway current density is decreased which in
turn largely decreases the avalanche generation rate.

\begin{figure}[htb!]
\begin{center}
\includegraphics[width=0.49\linewidth]{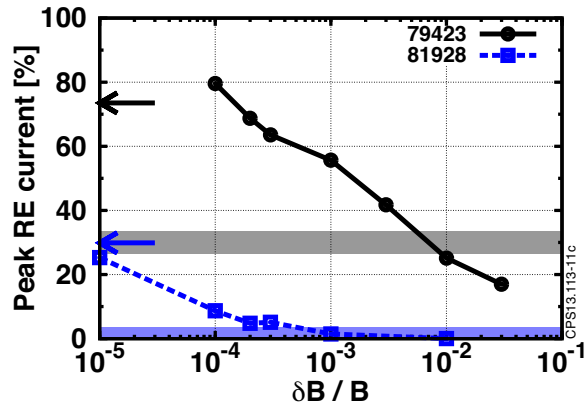}
\caption{\label{fig:dB2}{Peak runaway current normalized to the predisruption current as a function of $\delta$B/B for the two cases. Arrows mark the runaway fraction corresponding to $\delta B=0$. The grey and blue rectangles represent the experimentally measured range for the runaway current fraction.}}
\end{center}
\end{figure}

Although the plasma current evolution is well matched in both cases at
the same level of magnetic perturbation, the fraction of runaway
current is different in the two cases, just as was observed in the
experiments. Figures \ref{fig:dB1}c-d show that the runaway current
ramps up with time roughly as $1-\exp(-t/\tau_g)$, where $\tau_g$ is the loss time. If a $\delta B>0$
perturbation is present it also drops exponentially, but with a~different $\tau_l$ loss time constant. Figure \ref{fig:dB2} shows the \textit{peak} runaway
current normalized to the predisruption current as a function of
$\delta B/B$ for the two different shots. As a~comparison, grey
and blue rectangles show the experimentally measured range for the
runaway current fraction as measured in the \textit{plateau}. Arrows mark the
runaway fraction corresponding to $\delta B=0$ in the simulations (note the logarithmic $\delta B$ axis)
which is $\sim$73\% for the C wall case and $\sim$30 \% for the ILW
case. The maximum value of the runaway current drops exponentially as
a function of $\delta B$. Note that even relatively high RE currents
can be dropped to practically zero within a few tens of milliseconds
if the perturbation level is $\delta B/B$ > $2\times 10^{-4}$ (figure \ref{fig:dB1}c-d).  The
reduction is larger in shot \#81928 than in \#79423. The reason why the ILW case is more sensitive to the losses due
to magnetic perturbations than the C wall case is that in the C wall
case the Dreicer mechanism is significantly stronger (figure
\ref{fig:tdata79423-81928}a), generating a higher fraction of the runaway
current than in the ILW case (figure \ref{fig:tdata79423-81928}b). The Dreicer
generation has approximately an order of magnitude shorter characteristic rise time than the
avalanche mechanism, and therefore the losses due to radial diffusion
can more easily counteract the runaway growth if the Dreicer current
fraction is low. The impurity injection simulations have shown that
not only the runaway current but also the Dreicer fraction is
significantly lower for experimentally relevant argon and wall
impurity contents with the ILW (see figures \ref{fig:scan1_Irun} and
\ref{fig:scan1_Dfrac}.) Figure \ref{fig:dB2} shows that not only the
time evolution of the plasma current but also the order of magnitude of the runaway
current fraction is well matched with the experiments at the similar
level of magnetic perturbations, $\delta B/B$ > $2\times 10^{-4}$.

\section{Discussion and conclusions}\label{sec:conclusion}
As accurate knowledge about the realistic impurity levels is not
available, comparison between simulations and experiments need to be
based on reasonable assumptions about impurity content and rise
time. Our analysis focuses on the effect of temperature evolution,
impurity contents and magnetic perturbation levels on
runaway electron dynamics and current evolution. In general, the
results of the numerical simulations are in reasonable agreement with
the experimental observations.  The best agreement between simulation
and experimental observation in terms of current, radiation and
temperature evolution are reached at $\gtrsim$10\% background impurity
content, which is reasonable considering the relatively high \Zeff
even before argon injection. In terms of argon content the best agreement is found in the 20-30\% range that aligns well with the injected argon amounts at $\sim$30\% mixing.

Our results show that the differences between the C wall and ILW cases
are due to (1) the difference in the initial parameters, (2) the difference in the injected/mixed argon
amount, and (3) the different radiation characteristics of beryllium and carbon.
Since these three main differences between the discharges have a positive feedback
in terms of runaway generation, the effects of differences in the plasma parameters and injected argon amount are
enhanced by the presence of different wall (carbon or beryllium)
impurities.  Although the discharges were selected to be similar, in
the ILW case both the initial electron temperature and density are higher, which are exposed to a lower amount of injected argon.

This modelling shows that variations in the argon content in these shots
have a considerable effect on the runaway generation. The Dreicer fraction is
reduced by the presence of beryllium, but is almost unaffected by the
presence of carbon (at the same argon content). This results in a
lower Dreicer current generation in the ILW case compared with the C
wall case. The runaway population in the ILW case consists mostly of
slowly growing avalanche runaways and they are effectively transported
out from the plasma by a low level of magnetic perturbations or other losses. Note,
that the presence of beryllium is beneficial only if the amount of
argon is not too large. In our simulations the combination of 20\%
argon and 10\% beryllium content effectively reduced the Dreicer fraction of
the runaway current. Above 40-50\% argon content the differences due to plasma parameters and wall material are reduced and eventually vanish.
In view of the results of this paper, upcoming massive gas injection experiments with the ILW will most probably have to face with the reoccurance of runaways for the scenarios that produced runaways using MGI with the carbon wall.
Dedicated runaway experiments with the ILW on JET are necessary to be able to better estimate the runaway behaviour in ITER.

Further improvement of the numerical model would be to take into
account the spatio-temporal dependence of heat conductivity $\chi_e$
and the plasma movement towards the wall. Direct removal of runaways
via first wall scrape off is not expected in these cases, since the
runaway current channel width is smaller in the simulations than the
distance of the magnetic axis from the first wall in the corresponding
time instants of the discharges. However, the wall itself can
contribute to the removal of plasma current and energy as well as changes in the plasma inductance and thus may
influence the evolution of the other parameters.

\ack{This research was funded partly by the European Communities under the contract of Association between EURATOM and {\em Vetenskapsr{\aa}det} and was carried out within the framework of the European Fusion Development Agreement. The views and opinions expressed herein do not necessarily reflect those of the European Commission.
The authors are grateful to C. Sozzi, D. R\'efy, E. Delabie, M. Stamp, A. Boboc, G. Pokol and L. Capone for fruitful discussions.}

\section*{References}
\bibliographystyle{unsrt}
\bibliography{references}

\end{document}